\documentclass[useAMS,usenatbib]{mn2e}
\usepackage{aas_macros}
\usepackage{amsmath}
\usepackage{amssymb}
\usepackage{graphicx}
\usepackage{url}

\begin{document}

\title[Very massive stars in eclipsing binaries]{Identifying Stars of Mass $\mathbf{>150M_{\odot}}$ from Their Eclipse by a Binary Companion}
\author[T. Pan,  A. Loeb]{Tony Pan$^1$, Abraham Loeb$^1$\\
$^1$Harvard-Smithsonian Center for Astrophysics, 60 Garden Street, Cambridge, MA 02138, USA
}

\pagerange{\pageref{firstpage}--\pageref{lastpage}} \pubyear{2012}
\maketitle
\label{firstpage}

\begin{abstract}

We examine the possibility that very massive stars greatly exceeding
the commonly adopted stellar mass limit of $150 M_{\odot}$ may be
present in young star clusters in the local universe.  We identify ten
candidate clusters, some of which may host stars with masses up to
$600 M_{\odot}$ formed via runaway collisions.  We estimate the
probabilities of these very massive stars being in eclipsing binaries
to be $\gtrsim 30\%$.  Although most of these systems cannot be
resolved at present, their transits can be detected at distances of 3
Mpc even under the contamination of the background cluster light, due
to the large associated luminosities $\sim 10^7 L_{\odot}$ and mean
transit depths of $\sim 10^6 L_{\odot}$.  Discovery of very massive
eclipsing binaries would flag possible progenitors of pair-instability
supernovae and intermediate-mass black holes.

\end{abstract}

\label{lastpage}


\section{Introduction}

Many observations support the statistical argument that the upper
limit to initial stellar masses is $\sim 150 M_{\odot}$ for Pop II/I
stars \citep{Figer2005, Zinnecker2007}.  However, this common notion is
challenged by the recent spectroscopic analyses of
\citet{Crowther2010}, in which star clusters NGC 3603 and R136 are
found to host several stars with initial masses above this limit,
including one star R136a1 with a current mass of $\sim 265 M_{\odot}$.  
Also, candidate pair-instability supernovae, which
require progenitors with masses above $200 M_{\odot}$, have been
observed in the low redshift universe \citep{Gal-Yam2009}.  Therefore,
it is worth exploring methods to confirm the existence of a very
massive star (VMS), defined here as a star with a stellar mass
significantly greater than the stellar mass limit, i.e. $M \gtrsim 200 M_{\odot}$.

Unless the VMS is very close by, it is extremely difficult to
spatially resolve the VMS from stars in its vicinity.  Indeed, the
central component of R136 was once thought to be an extremely massive
$\gtrsim 10^3 M_{\odot}$ star \citep{Cassinelli1981}, before
\citet{Weigelt1985} resolved it as a dense star cluster via speckle
interferometry.  As for spectroscopic measurements, verification of a
single VMS is further complicated by the fact that the effective
temperature $T_{eff}$ of Pop I stars above $10^2 M_{\odot}$ depends
very weakly on mass, with $\log (T_{eff}/{\rm K})\approx 4.7$--$4.8$
\citep{Bromm2001} for stars between $10^2$--$10^3 M_{\odot}$.
Moreover, a hot evolved star with an initial mass below $10^2 M_{\odot}$ 
can nevertheless reach these temperatures in its post
main-sequence evolution and mimic a VMS.

The most accurate method of constraining the
stellar masses of distant stars is by measuring the radial velocity
and light curves of the star in an eclipsing binary
\citep{Bonanos2009, Torres2010}.  The light curve provides a wealth of
information about the binary, including its orbital period,
inclination, eccentricity, as well as the fractional radii and flux
ratio of the binary members.  The radial velocities found from a
double-lined spectroscopic binary further provide the mass ratio of
the binary.  With the above information, the individual masses of each
star in the binary can be calculated via Kepler's third law.  Searches
for massive eclipsing binaries in star clusters within our own Galaxy
are already underway \citep{Koumpia2011}, and techniques
have been suggested for binary searches in other galaxies \citep{Bonanos2012}.

In this {\it Letter}, we estimate the masses and properties of VMSs
that may have formed via collision runaways in a number of very young,
dense, and massive star clusters in the local universe.  We calculate
the probability of these VMSs to be in eclipsing binaries, and find
their expected transit depths and observability.

\section{Very Massive Stars}

Shortly after a dense star cluster forms, its most massive
constituents sink to the center via dynamic friction and form a
central subsystem of massive stars.  In sufficiently dense
environments, these massive stars may undergo runaway collisions and
merge into a single VMS \citep{Gurkan2004, Freitag2006}, possibly up
to $\sim 10^3 M_{\odot}$.  \citet{PortegiesZwart2006} gives a fitting
formula for the stellar mass $m_r$ of the final runaway product,
calibrated by N-body simulations for Salpeter-like mass functions:
\begin{equation}
m_r \sim 0.01 \: M_C \: \left(1+\frac{t_{rh}}{100 {\rm Myr}}\right)^{-\frac{1}{2}},
\label{RunawayMassEq}
\end{equation}
where $t_{rh}$ is the relaxation time,
\begin{equation}
t_{rh}	\approx 200 \: {\rm Myr} \left(\frac{r_{vir}}{1 {\rm pc}}\right)^{\frac{3}{2}}\left(\frac{ M_C }{10^6 M_{\odot}}\right)^{\frac{1}{2}}\frac{\langle m \rangle}{M_{\odot}}.
\label{RelaxationTimeEq}
\end{equation}
Here $M_C$ is the cluster mass, $r_{vir}$ is its virial radius, and
$\langle m \rangle \approx 0.5 M_{\odot}$ is the average stellar
mass.

Using the compilation of stars clusters in the local universe and
their properties from \citet{PortegiesZwart2010}, we have listed in
Table \ref{TableOfClusterAndVMSMass} several young, dense star
clusters that may host a runaway collision product of mass $\gtrsim 200 M_{\odot}$ 
which may have not yet ended its life as a star.  We
restrict our sample to clusters with mean determined ages younger than
3.5 Myr.  This may already be insufficiently selective, as stars born
with masses $\gtrsim 200 M_{\odot}$ are expected to have lifetimes of
only 2-3 Myr \citep{Yungelson2008}; however, in the runaway collision
scenario, the VMS builds up its extraordinary mass via mergers over
$\sim 1-2$ Myr, and therefore its host cluster may have an age
exceeding the 2-3 Myr limit.  Of course, these observed cluster
properties should not be taken as certain; for example,
\citet{Ubeda2007} find the ages of NGC 4214 I-A and I-B to be $\sim 4$--$5$ Myr, 
likely too old for a VMS to be present.  Conversely,
there may be candidate clusters with VMSs that we have missed.  The
predicted runaway masses are only approximate, but give a sense of the
mass range of VMSs that may lurk at the center of these very young and
dense clusters.

Alternatively, if feedback effects are moderate, it may be possible
for a protostar to grow without a fixed mass limit via mergers or via
the accretion of extremely dense gas.  In this case, the mass of the
most massive star $m_u$ formed in a molecular cloud scales with the
mass of that cloud, and thus will be correlated with the mass of its
eventual host cluster \citep{Larson1982, Larson2003, Weidner2010}:
\begin{equation}
m_u \approx 1.2 \: {M_C}^{0.45}.
\label{UpperMassToClusterMassEq}
\end{equation}
If the above relationship is valid for cluster masses $>5\times 10^4 M_{\odot}$, 
VMSs will not be restricted to dense clusters, since a
collision runaway is no longer necessary for achieving masses 
$\gtrsim 150M_{\odot}$ (see Table \ref{TableOfClusterAndVMSMass}).

\begin{table*}
\begin{minipage}{170mm}
\begin{center}
\caption{Possible very massive stars in star clusters and their
eclipse probabilities.  The predicted runaway collision product mass
$m_r$ is calculated from equation (\ref{RunawayMassEq}).  Another
possible VMS stellar mass $m_u$ is calculated via the relationship
between the cluster mass and its most massive star in equation
(\ref{UpperMassToClusterMassEq}).  All masses are in units of
$M_{\odot}$, the cluster age is measured in Myr, and the virial radius
$r_{vir}$ is in units of pc.  If we optimistically choose the largest
mass of $m_r$ and $m_u$ for the primary mass $M_1$, we can calculate
its luminosity $L_1$ (in $L_{\odot}$) and radius $R_1$ (in
$R_{\odot}$) using the models of \citet{Bromm2001}, assuming a
characteristic stellar metallicity $(Z/Z_{\odot})=0.3$.  We calculate
the eclipsing probability $P_e$ assuming that the companion is a B0
star, although the result is weakly sensitive to the companion
mass. For generality, the expected transit depth
$\langle\delta\rangle$ is averaged over a uniform distribution in the
binary mass ratio $q$, up to a companion mass of $10^2 M_{\odot}$,
assuming non-grazing orbits, i.e. $\delta\approx (R_2/R_1)^2$.  For
all VMS candidates below, the expected dip in luminosity from the
eclipse is $\sim 10^6 L_{\odot}$.}

\begin{tabular}{ | l | l | l | l | l | l | l | l | l | l | l | l |}
\hline
Galaxy 	& Name		& Ref		& Age & $\log{M_C}$	& $r_{vir}$ & $m_r$		& $m_u$		& $L_1$	& $R_1$	& $P_e$	& $\langle\delta\rangle$ \\ \hline
Milky Way& Arches		& 1		& 2.0	& 4.30			& 0.68		& \bf{192}	& 103			& 5e6		& 44		& 39\%	& 16\%\\
LMC 		& R136		& 2,3,4	& 3.0	& 4.78			& 2.89 		& \bf{406}	& 170			& 1e7		& 61		& 36\%	& 8\%	\\
SMC 		& NGC 346	& 5		& 3.0	& 5.60			& 15.28		& \bf{640}	& 397			& 2e7		& 76		& 34\%	& 5\%	\\
M33 		& NGC 604  	& 6		& 3.5	& 5.00	  		& 48.21		& 97			& \bf{213}	& 6e6		& 46		& 38\%	& 15\%\\
NGC 1569 & C 			& 6		& 3.0	& 5.16			& 4.50  		& \bf{672}	& 252			& 2e7		& 77		& 34\%	& 5\%	\\
NGC 4214 & I-A 		& 6		& 3.5	& 5.44	  		& 28.69		& \bf{305}	& 337			& 1e7		& 56		& 36\%	& 10\%\\
NGC 4214 & I-B 		& 6		& 3.5	& 5.40	  		& 9.85		& \bf{619}	& 323			& 2e7		& 74		& 34\%	& 6\%	\\
NGC 4214 & II-C 		& 6		& 2.0	& 4.86			& 23.43		& 129			& \bf{185}	& 5e6		& 43		& 39\%	& 17\%\\
NGC 4449 & N-2 		& 6		& 3.0	& 5.00			& 3.57  		& \bf{565}	& 213			& 2e7		& 71		& 35\%	& 6\%	\\
NGC 5253 & IV 			& 6		& 3.5	& 4.72			& 5.26  		& \bf{271}	& 160			& 8e6		& 51		& 37\%	& 12\%\\ \hline
\end{tabular}
\\
(1) \citet{Figer1999}; (2) \citet{Hunter1995}; (3) \citet{Mackey2003}; (4) \citet{Andersen2009}; (5) \citet{Sabbi2008}; (6) \citet{Maiz-Apellaniz2001}. 

\label{TableOfClusterAndVMSMass}
\end{center}
\end{minipage}
\end{table*}

\section{Eclipse Probability}

The fraction of massive O-type stars in binaries $f_b$ is observed to
be extremely high $> 70\%$ \citep{Chini2012}, and approaches 100\% in
some environments \citep{Mason2009, Bosch2009}.  Although there is no
related observational data on VMSs, numerical simulations indicate
that the collision runaway product in young, dense star clusters is
generally accompanied by a companion star (Portegies Zwart, private
communication).

As the period distribution for our hypothetical VMS binaries is
unknown, we assume their periods share the same cumulative
distribution function (CDF) as the periods of massive binaries
determined from observations.  The CDF of the orbital period ($p$, in
days) for massive binaries follows a `broken' \"Opik law, i.e. a
bi-uniform distribution in $\log{p}$, with the break at $p=10$
\citep{Sana2011}.  There is an overabundance of short period binaries,
with 50\% to 60\% of binaries having periods less than 10 days.  The
corresponding probability distribution function $PDF(p)$ of the
orbital period is:
\[
  PDF(p) = \frac{1}{\ln{10}} \times \left\{
  \begin{array}{l l}
    \frac{5}{7p}, & \quad \textrm{for $10^{0.3} \leq p \leq 10$ }\\\\
    \frac{1}{5p}, & \quad \textrm{for $10 < p \leq 10^{3.5}$ ,}\\
  \end{array} \right.
\] 
with the normalization $\int PDF(p) dp = 1$.

By integrating over uniformly distributed inclinations, it is easy to
show that the eclipsing probability of a binary system at any depth is
$P_e(a) = \frac{R_t}{a}$, where $R_t=R_1+R_2$ is the sum of the radii
of both components in the binary, and $a$ is the orbital distance.
From Kepler's third law, we can express the eclipsing probability as a
function of $p$ instead:
\begin{equation}
P_e(p) = R_t\left(\frac{2\pi}{p}\right)^{\frac{2}{3}}(G M_t)^{-\frac{1}{3}},
\end{equation}
where $M_t=M_1+M_2$ is the total system mass.  Therefore, integrating
over the period distribution, the probability that a massive binary
will be an eclipsing binary to an observer on Earth is
\begin{eqnarray}
\nonumber
P_{e} 	&=&	\int P_e(p) \: PDF(p) \: dp  \\
					&\approx &	0.053 \left[\frac{R_t}{R_{\odot}}\right] \left[\frac{M_t}{M_{\odot}}\right]^{-\frac{1}{3}}.
\label{FinalEclipseProbability}
\end{eqnarray}
For convenience, we ignore any effects of eccentricity; tidal
evolution will rapidly circularize the orbit for binaries with
periods below $p=10$ days, which account for 88\% of the above eclipsing systems.
Dynamical effects would harden a wide-separation massive binary system
in the core of a dense cluster on a timescale much shorter than 1 Myr.
Since three-body interactions tend to eject the lightest star, the
companion to the VMS will likely be a massive star, though not as
massive as the runaway product.

The large radii of VMSs coupled with their high binary fraction (and
short period binaries being common), imply significant eclipsing
probabilities for VMSs.  Using R136a1 as an example of the primary
star, with a radius $\sim 35 R_{\odot}$,
and a secondary Sun-like star, the eclipsing probability is $29\%$,
while for a more massive secondary star more common in the core of a
young massive star cluster, e.g. a B0 star of mass $\sim 18 M_{\odot}$
and radius $\sim 7 R_{\odot}$, the eclipse probability is $34\%$.
Note that the eclipsing binary probability in equation
(\ref{FinalEclipseProbability}) is not sensitive to the secondary star
parameters, as long as its radius is small relative to the primary.

Assuming a companion B0 star, we list the eclipsing binary
probabilities for our candidate VMSs in Table
\ref{TableOfClusterAndVMSMass}, calculated from equation
(\ref{FinalEclipseProbability}), except that we limit the integration
over $p$ to periods corresponding to orbital distances exceeding both
the radius of the VMS and the Roche limit for the companion.  This
restriction reduces $P_e$, and leads to the larger VMSs having
slightly smaller eclipsing probabilities; nevertheless, the eclipsing
probabilities for all VMS candidates exceed 1/3.

\section{Observability of Transit}

VMSs have spectacular luminosities in the range of $10^7 L_{\odot}$;
for example, R136a1 is observed to have $\sim 8.7\times 10^6 L_{\odot}$.  
Even at a distance of 3 Mpc -- roughly the distance of
the farthest host galaxy in Table \ref{TableOfClusterAndVMSMass} -- a
star like R136a1 would still have an apparent bolometric magnitude of 14.8.  
However, VMSs with $T_{eff}\sim 5\times 10^4$ K emit primarily in the ultraviolet,
requiring bolometric corrections of $BC\sim 4.6$.  Still, such a VMS
will be within the V-band limiting magnitude of ground-based 1-meter
telescopes.  For the VMS candidates in Table
\ref{TableOfClusterAndVMSMass}, with a hypothetical B0-star companion,
the transit depth exceeds $10^5 L_{\odot}$ in all cases, which at 3
Mpc is just within the single-visit limiting magnitudes of future
synoptic surveys such as
Pan-STARRS\footnote{\url{http://pan-starrs.ifa.hawaii.edu/public/}}
and the Large Synoptic Survey
Telescope\footnote{\url{http://www.lsst.org/lsst/}}.  Of course, given
the shortlist of host clusters in Table
\ref{TableOfClusterAndVMSMass}, one can use deep, targeted observations
of the individuals clusters with existing telescopes, instead of
uniform field surveys.

\begin{figure}
\begin{center}
\includegraphics[width=0.95\columnwidth]{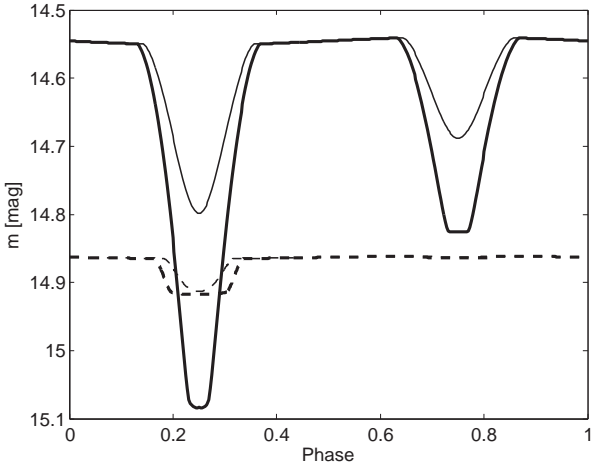}
\caption{Example light curves for a VMS eclipsing binary.  The primary
has parameters similar to R136a1, while the secondary is either a $18
M_{\odot}$ (dashed line) or $100 M_{\odot}$ (straight line) star, with
appropriate radii and luminosities.  The apparent magnitude $m$
(bolometric) is plotted for these systems at 3 Mpc.  The thick and
thin lines correspond to inclinations of $90^{\circ}$ and
$70^{\circ}$, respectively; the period is 5 days in both cases.
Reflections and limb-darkening using the model of
\citet{Diaz-Cordoves1995} are taken into account, but ellipsoidal
variation is ignored.}
\label{PlotLightCurve}
\end{center}
\end{figure}

However, in massive binaries, the mass ratio between the primary and
secondary star $q=M_2/M_1$ is observed to have a flat distribution
\citep{Sana2011}.  Unlike the transit probability, the transit depth
is very sensitive to the companion star radii, so using a B0 star as
the companion may be overly conservative.  Since only one VMS is
expected to form in the collision runaway scenario, here we assume the
distribution of companion star masses is uniform between 1 to 100
$M_{\odot}$.  Using typical mass-radius relationships, we show in
Table \ref{TableOfClusterAndVMSMass} the expected transit depth
$\langle\delta\rangle$ integrated over the range of companion star
radii.  Figure \ref{PlotLightCurve} illustrates sample light curves
for a VMS binary at 3 Mpc with different companion star masses and
radii at different inclinations.

For clusters outside the Milky Way and its satellites, it is currently
impossible to resolve a VMS from other massive stars in a dense
cluster core.  Hence, we consider the luminosity of the host cluster
as a contaminating third light source to the eclipsing binary light
curve.  If the VMS is present, it will contribute a significant
fraction of the bolometric luminosity of the cluster (at least $10\%$ and
exceeding $50\%$ in some cases), and an even larger fraction of the UV
flux.  The integration time $t$ needed to reach a target
signal-to-noise ratio $SNR$ for detecting a transit can be
approximated as:
\begin{eqnarray}
\nonumber
t &\approx& 6\: {\rm seconds} \times \left[\frac{L_C}{10^8 \: L_{\odot}}\right]^{-1} \left[\frac{d}{3 \: {\rm Mpc}}\right]^2 \\
\nonumber
&&\times \left[\frac{f_{band}}{0.2}\right]^{-1} 
\left[\frac{E_{band}}{10 \: {\rm eV}}\right] \left[ \frac{A}{4\times 10^4 \: {\rm cm}^2} \right]^{-1} \\
&&\times \left[\frac{SNR}{10}\right]^2 \left[\frac{f_{VMS}}{0.1}\right]^{-2} \left[\frac{\delta}{10\%}\right]^{-2}
\end{eqnarray}
where $L_C$ is the bolometric luminosity of the cluster, $d$ is the
distance to the cluster, $f_{band}$ is the fraction of total flux that is observed 
(due to the spectral energy distribution, filter bandpass, CCD response, atmospheric transmission etc.), 
$E_{band}$ is the characteristic observed photon energy, $A$ is the
collecting area of the telescope, $f_{VMS}$ is the fraction of total
observed flux from the VMS primary, and again $\delta$ is the transit
depth.

Note that the Hubble Space Telescope (HST) would collect $\gtrsim 10^4$ 
UV photons per second from a $10^7 L_{\odot}$ VMS even at a
distance of 3 Mpc, thus detecting a $\delta\sim 10\%$ transit depth at
$SNR=10$ in tens of seconds of integration time.  Obscuration by dust
along the line-of-sight may reduce the observed UV flux.  For V-band
observations, a very young $\sim 10^5 M_{\odot}$ cluster can be as
bright as $M_V\approx -12$, while a $300 M_{\odot}$ VMS will have
$M_V\approx -8$, i.e. the VMS will only contribute $f_{VMS}\sim 2.5\%$
of the cluster light in the visible band.  Nevertheless, a 2-meter
ground-based telescope will need less than an hour of integration time
to detect the transit, which is eminently feasible as the transit
duration $\tau\sim p({R_1}/{\pi a}) \propto p^{1/3}$ for a VMS
eclipsing binary will be $>10$ hours for all relevant orbital periods.

Other less massive eclipsing binaries in the host cluster will also
contaminate the light curve, but their transit depths will likely be
negligible compared to the VMS's luminosity.  Non-binary random
occultations of the central VMS can replicate a large transit depth,
but using a King model for the cluster density profile
\citep{King1966}, we find these events occur less than once every
$10^6$ years.

\section{Stellar Mass Determination}

The extraordinary luminosity of a VMS should allow its radial velocity
to be measured.  However, the mass ratio $q$, critical for model-independent
determination of the individual masses, can only be
found when the radial velocities are determined for \emph{both}
components of the binary.  Such double-lined spectroscopic binaries
are easily observable when the components have similar luminosities,
within a factor of 5 of each other \citep{Kallrath2009}.  As the
luminosity of massive stars near the Eddington limit scales with mass,
this criteria roughly corresponds to $q>0.2$, which for an uniform
distribution in $q \in (0,1)$ is quite likely to occur.

Nevertheless, if the companion is small, and only spectral lines from
the VMS are detected, then the mass ratio $q$ cannot be unambiguously
obtained.  Instead, the mass of the VMS can be expressed as a single
function of $q$:
\begin{equation}
M_1 = \frac{(1+q)^2}{q^3}\frac{1}{\sin^3{i}}f(M_1,M_2,i),
\label{M_1_single_line}
\end{equation}
where $f(M_1,M_2,i)$ is the \emph{mass function}, which can be
calculated using quantities derivable from the spectroscopy of a
single-lined spectroscopic binary, and the inclination $i$ is
derivable from the eclipsing binary light curve.  Unfortunately,
equation (\ref{M_1_single_line}) varies sharply as $\propto q^{-3}$
for $q\ll 1$. Since $q$ can be as small as $\sim 0.01$ for VMSs in
Table \ref{TableOfClusterAndVMSMass}, crude constraints on the mass
ratio, e.g. $q<0.2$ (when light from the secondary is not observed)
cannot establish tight minimum stellar mass constraints on the VMS
primary.

However, since a total eclipse $\delta \rightarrow 100\%$ is extremely
unlikely given the large radii of VMSs, if the mean value $\sim 10^6 L_{\odot}$ 
dip in the light curve is in fact observed, it will
immediately imply the existence of a star $\gtrsim 10^2 M_{\odot}$.
Hence, although sophisticated light-curve fitting with stellar models
would be required, eclipsing single-lined spectroscopic binaries still
offer an attractive avenue for inferring the presence of a VMS greatly
exceeding the 150 $M_{\odot}$ stellar mass limit.

\section{Discussion}

A search for periodic flux variations (as shown in Fig. 1) due to
transits of the VMS candidates in Table \ref{TableOfClusterAndVMSMass}
would be of considerable interest.  Although \citet{Crowther2010} made
robust arguments against R136a1 being a wide separation binary or an
equal-mass binary, this source could still involve a short-period,
unequal-mass binary system.  The Arches cluster is observed to have no
stars currently above the $150 M_{\odot}$ mass limit, but
\citet{Crowther2010} also found with contemporary stellar and
photometric results that the most luminous stars in the Arches cluster
had initial masses approaching $200 M_{\odot}$.

The radii of VMSs are dependent on their metallicities and rotation
\citep{Langer2007}.  
If the VMS radii in Table \ref{TableOfClusterAndVMSMass} were smaller
by $\sim 25\%$ (e.g. at much lower metallicities), all listed
eclipsing probabilities would still remain above 1/3, but the expected
transit depth would increase up to $\langle\delta\rangle\sim 20\%$.
As for the companion star, for most O stars, the point of unity
Thomson optical depth occurs close to the hydrostatic radius, but when
stellar mass loss exceeds $\sim 10^{-5} M_{\odot}$ yr$^{-1}$, the
photosphere $\tau\sim 1$ occurs in the wind itself, effectively
increasing the star's radius.  This occurs for Wolf-Rayet companions
\citep{Lamontagne1996} and for companions $\gtrsim 60 M_{\odot}$
\citep{Vink2000}, in which case our eclipse probabilities and transit
depths are too conservative.  

Binaries can be broadly classified into detached systems, where neither component
fills its Roche lobe, versus semi-detached or over-contact systems,
where at least one component exceeds its Roche lobe.  VMSs in detached
binaries have much more sharply defined eclipses, and more
importantly, they do not undergo mass transfer and lose mass to their
companion.  To find the probability that our VMS candidates in Table
\ref{TableOfClusterAndVMSMass} are detached eclipsing binaries, we
limit the integration in equation (\ref{FinalEclipseProbability}) to
periods $p \gtrsim 5$ days, corresponding to orbital distances where
the Roche lobe of the VMS is always greater than its radius
\citep{Eggleton1983}.  For our VMS candidates, the detached eclipsing
binary probability is $\approx 17 \%$, i.e. roughly half of all
eclipsing systems.

However, non-pristine massive stars can also lose mass via strong
winds driven by radiation pressure, with a mass loss rate increasing
with metallicity.  Post main-sequence VMSs can also lose mass
eruptively or via pulsational instabilities, although mass loss near
the end of the star's life (e.g. the pulsational pair-instability) is
not likely to change the observability of our VMS candidates.  Under
extraordinary mass loss via winds, \citet{Glebbeek2009} found the
highest mass attained by a collision runaway product to be $\sim 400
M_{\odot}$, although the star remained at this mass range for only
$\sim 0.2$ Myr.  On the contrary, \citet{Suzuki2007} found that
stellar mass loss does not inhibit the formation of a VMS of $\sim
10^3 M_{\odot}$.

If VMSs do in fact form via collision runaways in young, dense star
clusters, and retain sufficient masses at the end of their lives, they
may explode as pair-instability supernovae (PISNe)
\citep{Yungelson2008}. The creation rate of runaway products is in
fact consistent with the current observed PISN rate \citep{Pan2012}.
However, the most massive VMSs may collapse directly into an
intermediate mass black hole (IMBH) via the photodisintegration
instability \citep{Woosley2002}.  Tentative evidence has been claimed
for IMBHs at the center of old globular clusters
\citep{Lou2012}, and extragalactic ultraluminous x-rays sources
associated with young star clusters \citep{Ebisuzaki2001,
Farrell2009}.  The identification of VMSs that can serve as the
progenitors of PISNe and IMBHs will help move these extreme
astrophysical objects from the realm of speculation into reality.

\section*{Acknowledgments.}

We thank Paul Crowther, Dave Latham, Philip Myers, Guillermo Torres,
and Simon Portegies Zwart for helpful discussions.  We thank Eran Ofek
for his eclipsing binary
scripts\footnote{\url{http://wise-obs.tau.ac.il/~eran/matlab.html}},
which we built upon to generate our example light curves.  TP was
supported by the Hertz Foundation.  This work was supported in part by
NSF grant AST-0907890 and NASA grants NNX08AL43G and NNA09DB30A.

\bibliographystyle{mn2e}
\bibliography{references}

\begin{thebibliography}{44}
\expandafter\ifx\csname natexlab\endcsname\relax\def\natexlab#1{#1}\fi

\bibitem[{{Andersen} {et~al}\mbox{.}(2009){Andersen}, {Zinnecker}, {Moneti},
  {McCaughrean}, {Brandl}, {Brandner}, {Meylan}, \& {Hunter}}]{Andersen2009}
{Andersen} M., {Zinnecker} H., {Moneti} A., {McCaughrean} M.~J., {Brandl} B.,
  {Brandner} W., {Meylan} G., {Hunter} D., 2009, \apj, 707, 1347

\bibitem[{{Bonanos}(2009)}]{Bonanos2009}
{Bonanos} A.~Z., 2009, \apj, 691, 407

\bibitem[{{Bonanos}(2012)}]{Bonanos2012}
---, 2012, in IAU Symposium, Vol. 282, IAU Symposium, pp. 27--32

\bibitem[{{Bosch}, {Terlevich} \& {Terlevich}(2009){Bosch}, {Terlevich}, \&
  {Terlevich}}]{Bosch2009}
{Bosch} G., {Terlevich} E., {Terlevich} R., 2009, \aj, 137, 3437

\bibitem[{{Bromm}, {Kudritzki} \& {Loeb}(2001){Bromm}, {Kudritzki}, \&
  {Loeb}}]{Bromm2001}
{Bromm} V., {Kudritzki} R.~P., {Loeb} A., 2001, \apj, 552, 464

\bibitem[{{Cassinelli}, {Mathis} \& {Savage}(1981){Cassinelli}, {Mathis}, \&
  {Savage}}]{Cassinelli1981}
{Cassinelli} J.~P., {Mathis} J.~S., {Savage} B.~D., 1981, Science, 212, 1497

\bibitem[{{Chini} {et~al}\mbox{.}(2012){Chini}, {Hoffmeister}, {Nasseri},
  {Stahl}, \& {Zinnecker}}]{Chini2012}
{Chini} R., {Hoffmeister} V.~H., {Nasseri} A., {Stahl} O., {Zinnecker} H.,
  2012, ArXiv e-prints

\bibitem[{{Crowther} {et~al}\mbox{.}(2010){Crowther}, {Schnurr}, {Hirschi},
  {Yusof}, {Parker}, {Goodwin}, \& {Kassim}}]{Crowther2010}
{Crowther} P.~A., {Schnurr} O., {Hirschi} R., {Yusof} N., {Parker} R.~J.,
  {Goodwin} S.~P., {Kassim} H.~A., 2010, \mnras, 408, 731

\bibitem[{{Diaz-Cordoves}, {Claret} \& {Gimenez}(1995){Diaz-Cordoves},
  {Claret}, \& {Gimenez}}]{Diaz-Cordoves1995}
{Diaz-Cordoves} J., {Claret} A., {Gimenez} A., 1995, \aaps, 110, 329

\bibitem[{{Ebisuzaki} {et~al}\mbox{.}(2001){Ebisuzaki}, {Makino}, {Tsuru},
  {Funato}, {Portegies Zwart}, {Hut}, {McMillan}, {Matsushita}, {Matsumoto}, \&
  {Kawabe}}]{Ebisuzaki2001}
{Ebisuzaki} T. {et~al.}, 2001, \apjl, 562, L19

\bibitem[{{Eggleton}(1983)}]{Eggleton1983}
{Eggleton} P.~P., 1983, \apj, 268, 368

\bibitem[{{Farrell} {et~al}\mbox{.}(2009){Farrell}, {Webb}, {Barret}, {Godet},
  \& {Rodrigues}}]{Farrell2009}
{Farrell} S.~A., {Webb} N.~A., {Barret} D., {Godet} O., {Rodrigues} J.~M.,
  2009, \nat, 460, 73

\bibitem[{{Figer}(2005)}]{Figer2005}
{Figer} D.~F., 2005, \nat, 434, 192

\bibitem[{{Figer}, {McLean} \& {Morris}(1999){Figer}, {McLean}, \&
  {Morris}}]{Figer1999}
{Figer} D.~F., {McLean} I.~S., {Morris} M., 1999, \apj, 514, 202

\bibitem[{{Freitag}, {G{\"u}rkan} \& {Rasio}(2006){Freitag}, {G{\"u}rkan}, \&
  {Rasio}}]{Freitag2006}
{Freitag} M., {G{\"u}rkan} M.~A., {Rasio} F.~A., 2006, \mnras, 368, 141

\bibitem[{{Gal-Yam} {et~al}\mbox{.}(2009){Gal-Yam}, {Mazzali}, {Ofek},
  {Nugent}, {Kulkarni}, {Kasliwal}, {Quimby}, {Filippenko}, {Cenko},
  {Chornock}, {Waldman}, {Kasen}, {Sullivan}, {Beshore}, {Drake}, {Thomas},
  {Bloom}, {Poznanski}, {Miller}, {Foley}, {Silverman}, {Arcavi}, {Ellis}, \&
  {Deng}}]{Gal-Yam2009}
{Gal-Yam} A. {et~al.}, 2009, \nat, 462, 624

\bibitem[{{Glebbeek} {et~al}\mbox{.}(2009){Glebbeek}, {Gaburov}, {de Mink},
  {Pols}, \& {Portegies Zwart}}]{Glebbeek2009}
{Glebbeek} E., {Gaburov} E., {de Mink} S.~E., {Pols} O.~R., {Portegies Zwart}
  S.~F., 2009, \aap, 497, 255

\bibitem[{{G{\"u}rkan}, {Freitag} \& {Rasio}(2004){G{\"u}rkan}, {Freitag}, \&
  {Rasio}}]{Gurkan2004}
{G{\"u}rkan} M.~A., {Freitag} M., {Rasio} F.~A., 2004, \apj, 604, 632

\bibitem[{{Hunter} {et~al}\mbox{.}(1995){Hunter}, {Shaya}, {Holtzman}, {Light},
  {O'Neil}, \& {Lynds}}]{Hunter1995}
{Hunter} D.~A., {Shaya} E.~J., {Holtzman} J.~A., {Light} R.~M., {O'Neil}, Jr.
  E.~J., {Lynds} R., 1995, \apj, 448, 179

\bibitem[{{Kallrath} \& {Milone}(2009)}]{Kallrath2009}
{Kallrath} J., {Milone} E.~F., 2009, {Eclipsing Binary Stars: Modeling and
  Analysis}

\bibitem[{{King}(1966)}]{King1966}
{King} I.~R., 1966, \aj, 71, 64

\bibitem[{{Koumpia} \& {Bonanos}(2011)}]{Koumpia2011}
{Koumpia} E., {Bonanos} A.~Z., 2011, in IAU Symposium, Vol. 272, IAU Symposium,
  {Neiner} C., {Wade} G., {Meynet} G., {Peters} G., eds., pp. 515--516

\bibitem[{{Lamontagne} {et~al}\mbox{.}(1996){Lamontagne}, {Moffat}, {Drissen},
  {Robert}, \& {Matthews}}]{Lamontagne1996}
{Lamontagne} R., {Moffat} A.~F.~J., {Drissen} L., {Robert} C., {Matthews}
  J.~M., 1996, \aj, 112, 2227

\bibitem[{{Langer} {et~al}\mbox{.}(2007){Langer}, {Norman}, {de Koter}, {Vink},
  {Cantiello}, \& {Yoon}}]{Langer2007}
{Langer} N., {Norman} C.~A., {de Koter} A., {Vink} J.~S., {Cantiello} M.,
  {Yoon} S.-C., 2007, \aap, 475, L19

\bibitem[{{Larson}(1982)}]{Larson1982}
{Larson} R.~B., 1982, \mnras, 200, 159

\bibitem[{{Larson}(2003)}]{Larson2003}
---, 2003, Reports on Progress in Physics, 66, 1651

\bibitem[{{Lou} \& {Wu}(2012)}]{Lou2012}
{Lou} Y.-Q., {Wu} Y., 2012, \mnras, 422, L28

\bibitem[{{Mackey} \& {Gilmore}(2003)}]{Mackey2003}
{Mackey} A.~D., {Gilmore} G.~F., 2003, \mnras, 338, 85

\bibitem[{{Ma{\'{\i}}z-Apell{\'a}niz}(2001)}]{Maiz-Apellaniz2001}
{Ma{\'{\i}}z-Apell{\'a}niz} J., 2001, \apj, 563, 151

\bibitem[{{Mason} {et~al}\mbox{.}(2009){Mason}, {Hartkopf}, {Gies}, {Henry}, \&
  {Helsel}}]{Mason2009}
{Mason} B.~D., {Hartkopf} W.~I., {Gies} D.~R., {Henry} T.~J., {Helsel} J.~W.,
  2009, \aj, 137, 3358

\bibitem[{{Pan}, {Loeb} \& {Kasen}(2012){Pan}, {Loeb}, \& {Kasen}}]{Pan2012}
{Pan} T., {Loeb} A., {Kasen} D., 2012, \mnras, 2979

\bibitem[{{Portegies Zwart} {et~al}\mbox{.}(2006){Portegies Zwart},
  {Baumgardt}, {McMillan}, {Makino}, {Hut}, \&
  {Ebisuzaki}}]{PortegiesZwart2006}
{Portegies Zwart} S.~F., {Baumgardt} H., {McMillan} S.~L.~W., {Makino} J.,
  {Hut} P., {Ebisuzaki} T., 2006, \apj, 641, 319

\bibitem[{{Portegies Zwart}, {McMillan} \& {Gieles}(2010){Portegies Zwart},
  {McMillan}, \& {Gieles}}]{PortegiesZwart2010}
{Portegies Zwart} S.~F., {McMillan} S.~L.~W., {Gieles} M., 2010, \araa, 48, 431

\bibitem[{{Sabbi} {et~al}\mbox{.}(2008){Sabbi}, {Sirianni}, {Nota}, {Tosi},
  {Gallagher}, {Smith}, {Angeretti}, {Meixner}, {Oey}, {Walterbos}, \&
  {Pasquali}}]{Sabbi2008}
{Sabbi} E. {et~al.}, 2008, \aj, 135, 173

\bibitem[{{Sana} \& {Evans}(2011)}]{Sana2011}
{Sana} H., {Evans} C.~J., 2011, in IAU Symposium, Vol. 272, IAU Symposium,
  {C.~Neiner, G.~Wade, G.~Meynet, \& G.~Peters}, ed., pp. 474--485

\bibitem[{{Suzuki} {et~al}\mbox{.}(2007){Suzuki}, {Nakasato}, {Baumgardt},
  {Ibukiyama}, {Makino}, \& {Ebisuzaki}}]{Suzuki2007}
{Suzuki} T.~K., {Nakasato} N., {Baumgardt} H., {Ibukiyama} A., {Makino} J.,
  {Ebisuzaki} T., 2007, \apj, 668, 435

\bibitem[{{Torres}, {Andersen} \& {Gim{\'e}nez}(2010){Torres}, {Andersen}, \&
  {Gim{\'e}nez}}]{Torres2010}
{Torres} G., {Andersen} J., {Gim{\'e}nez} A., 2010, \aapr, 18, 67

\bibitem[{{{\'U}beda}, {Ma{\'{\i}}z-Apell{\'a}niz} \&
  {MacKenty}(2007){{\'U}beda}, {Ma{\'{\i}}z-Apell{\'a}niz}, \&
  {MacKenty}}]{Ubeda2007}
{{\'U}beda} L., {Ma{\'{\i}}z-Apell{\'a}niz} J., {MacKenty} J.~W., 2007, \aj,
  133, 932

\bibitem[{{Vink}, {de Koter} \& {Lamers}(2000){Vink}, {de Koter}, \&
  {Lamers}}]{Vink2000}
{Vink} J.~S., {de Koter} A., {Lamers} H.~J.~G.~L.~M., 2000, \aap, 362, 295

\bibitem[{{Weidner}, {Kroupa} \& {Bonnell}(2010){Weidner}, {Kroupa}, \&
  {Bonnell}}]{Weidner2010}
{Weidner} C., {Kroupa} P., {Bonnell} I.~A.~D., 2010, \mnras, 401, 275

\bibitem[{{Weigelt} \& {Baier}(1985)}]{Weigelt1985}
{Weigelt} G., {Baier} G., 1985, \aap, 150, L18

\bibitem[{{Woosley}, {Heger} \& {Weaver}(2002){Woosley}, {Heger}, \&
  {Weaver}}]{Woosley2002}
{Woosley} S.~E., {Heger} A., {Weaver} T.~A., 2002, Reviews of Modern Physics,
  74, 1015

\bibitem[{{Yungelson} {et~al}\mbox{.}(2008){Yungelson}, {van den Heuvel},
  {Vink}, {Portegies Zwart}, \& {de Koter}}]{Yungelson2008}
{Yungelson} L.~R., {van den Heuvel} E.~P.~J., {Vink} J.~S., {Portegies Zwart}
  S.~F., {de Koter} A., 2008, \aap, 477, 223

\bibitem[{{Zinnecker} \& {Yorke}(2007)}]{Zinnecker2007}
{Zinnecker} H., {Yorke} H.~W., 2007, \araa, 45, 481

\end{thebibliography}

\end{document}